\documentclass[aps,reprint,article,prl]{revtex4-2}
\usepackage{amsmath}
\usepackage{dcolumn}
\usepackage{graphicx}
\usepackage{float}
\usepackage{color}
\usepackage{booktabs}
\usepackage{multirow}
\usepackage{hyperref}
\hypersetup{
	colorlinks=true,
	linkcolor=blue,
	filecolor=blue,      
	urlcolor=blue,
	citecolor=blue,
}

\begin{document}

\title{Distinct Modulation Behavior of Superconducting Coherence Peaks Associated with Sign-Reversal Gaps in FeTe$_{0.55}$Se$_{0.45}$}

\author{Zhiyong Hou,$^{*}$ Zhiyuan Shang,$^{*}$ Wen Duan, Wei Xie, Huan Yang,$^{\dag}$ and Hai-Hu Wen$^{\S}$}

\affiliation{National Laboratory of Solid State Microstructures and Department of Physics, Collaborative Innovation Center of Advanced Microstructures, Nanjing University, Nanjing 210093, China}

\begin{abstract}
 Using high-resolution scanning tunneling microscopy, we reveal two distinct types of superconducting (SC) gap modulations in bulk superconductor FeTe$_{0.55}$Se$_{0.45}$. By analyzing the phase relation between modulations at positive and negative bias, we identify in-phase (particle–hole asymmetric) and anti-phase (particle–hole symmetric) oscillations, corresponding to sign-reversing and sign-preserving scattering processes, respectively. The observed features are consistent with predictions from pair-breaking scattering interference (PBSI) theory and are distinguishable from other alternative mechanisms such as pair density waves. Our results provide compelling evidence that PBSI is the dominant mechanism behind the SC gap modulations in FeTe$_{0.55}$Se$_{0.45}$, offering new insights into the role of impurity scattering in iron-based superconductors.
\end{abstract}

\maketitle

Superconducting (SC) states with spatially modulated order parameters, such as the pair density wave (PDW), has become a pivotal topic in superconductivity. However, the mechanism behind SC gap modulation remains controversial. The PDW state is an interesting SC phase, and the SC order parameter has a spatial modulation with an averaged value of nearly zero \cite{1,2}. The PDW phase has attracted intense interest as a possible intertwined or competing phase within the SC state. Experimentally, the PDW order was observed in many kinds of superconductors, including cuprate \cite{1,2,3,4,5}, iron-based \cite{6,7}, transition metal dichalcogenides \cite{8,9}, kagome metals \cite{10,11}, and heavy fermion superconductors \cite{12}. In these measurements, the obtained local SC gap or local superfluid density has a spatial oscillation with a commensurate or incommensurate period compared to the lattice constant of the sample. Such an interesting SC state provides a new perspective for the study of unconventional superconductivity. Recently, the pair density modulations (PDM) were observed in some iron chalcogenides in the form of thin flakes \cite{13} or monolayer films \cite{14,15}. In these works, the SC gap has a modulation with the period of the lattice constant on the surface, and it is argued that the PDM state is driven by the interplay of glide-mirror symmetry breaking and nematic superconductivity \cite{13,16}.

The PDM has a period of the lattice constant, and the atomic lattice can obviously induce a spatial oscillation of local density of states (LDOS). Meanwhile, it should be noted that impurities can also induce quasiparticle interference (QPI) on the surface and result in spatial oscillations of LDOS, and the oscillation period in real space corresponds to scattering vectors connecting the hot spots of the Fermi surfaces \cite{17}. In a superconductor, the QPI obtained at an energy within or near the gap energy may carry information about the SC gap, including the gap anisotropy \cite{18,19} and even the gap signs \cite{20,21,22,23,24,25,26,27,28}. Recently, a theoretical model was proposed, which suggests that the SC gap modulation can be induced by the pair-breaking scattering interference (PBSI) \cite{29}, and the scattering vectors connecting hotspots on Fermi surface(s) with the same (or opposite) SC gap signs can produce anti-phase (or in-phase) modulations of coherence peak energies at positive and negative sides. Therefore, it is interesting to check the situation in a superconductor with both sign-preserving and sign-reversing gaps.

In this study, we perform high-resolution scanning tunneling microscopy/spectroscopy (STM/STS) measurements on a sub-atomic scale in the single crystals of iron-based superconductor FeTe$_{0.55}$Se$_{0.45}$ with the $s^\pm$ pairing manner. We observe two distinct types of coherence peak modulations in real space coexisting in a superconductor with periods corresponding to different inter-pocket scattering wave vectors. By analyzing the phase relation between modulations at positive and negative energies, we identify in-phase (particle-hole asymmetric) and anti-phase (particle-hole symmetric) oscillations of coherence peak energies, corresponding to gap-sign-reversing and gap-sign-preserving pair-breaking scattering processes, respectively. Our results provide an important view of SC gap modulation in superconductors.

\begin{figure}[tbp]
\centering
\includegraphics[width=8cm]{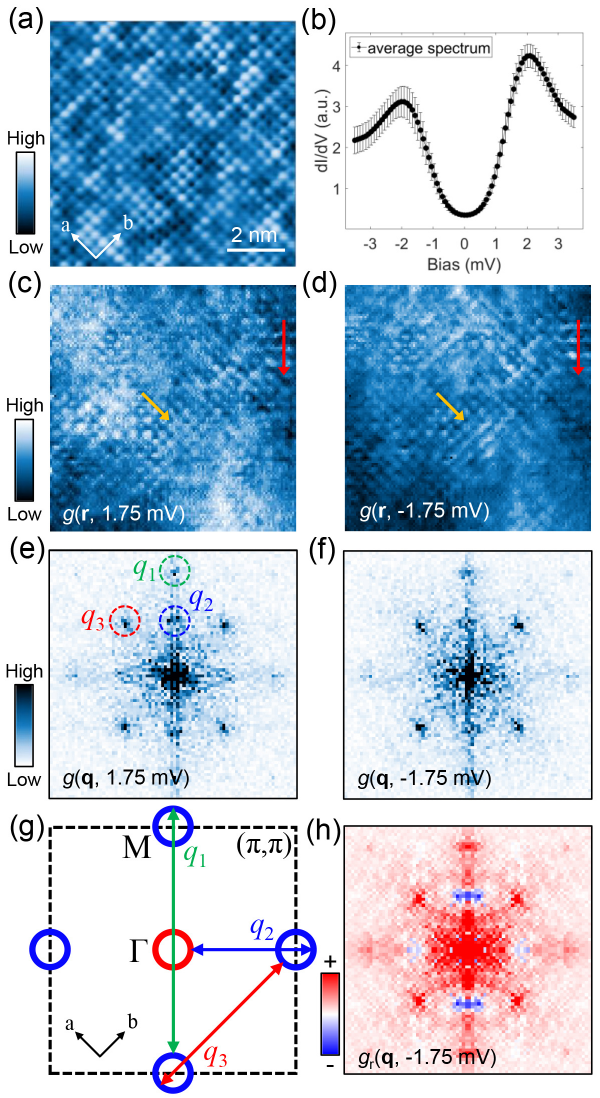}
\caption{(a) Atomically resolved topography measured on FeTe$_{0.55}$Se$_{0.45}$. (b) Average tunneling spectrum of 112×112 spectra measured in the area of (a). (c),(d) QPI images $g(\textbf{r}, E)$ in the real space measured at +1.75 and -1.75 mV, respectively. (e),(f) FT-QPI patterns $|g(\textbf{q}, E)|$ obtained from the Fourier transform results of (c) and (d), respectively. (g) Schematic image of the Fermi surfaces in the unfolded Brillouin zone. The hole pocket is centered at $\Gamma$ point, while the electron pockets are around M points. (h) PR-QPI pattern $g_r(\textbf{q}, E)$ obtained by the DBS-QPI technique. The pattern is fourfold-symmetrized.
} \label{fig1}
\end{figure}

Figure \ref{fig1}(a) shows a representative topographic image of a single crystal of FeTe$_{0.55}$Se$_{0.45}$ after cleavage. One can see that the atoms on the surface construct a square lattice with a lattice constant $a_0\approx3.8\mathrm{~\r{A}}$, and the value is about $\sqrt{2}$ times of the lattice constant $a_{Fe-Fe}$ of the Fe lattice underneath. The brighter atoms on the surface are Te atoms because they have a larger atomic size, while the darker ones are Se atoms. The topographic image is consistent with previous reports \cite{20,27,30,31,32}. It should be noted that the tunneling spectra are very inhomogeneous on bulk FeTe$_{0.55}$Se$_{0.45}$ \cite{33,34}. Therefore, we plot in Fig. \ref{fig1}(b) the average spectrum of 112×112 measured spectra obtained in Fig. \ref{fig1}(a) in order to show the superconductive feature of the sample. The average spectrum shows coherence peaks at about $\pm$2.0 mV, which suggests an average SC gap or the gap maximum of about 2 meV. Meanwhile, there is a finite differential conductance at zero bias, which may be due to some scattering on the top surface after the average effect.

The quasiparticle interference (QPI) image can provide fruitful information about the electronic state of the material \cite{17} and even the SC gap \cite{18,19,20,21,22,23,24,25,26,27,28}. In Figs. \ref{fig1}(c) and \ref{fig1}(d), we plot the QPI images measured with bias-voltage values of $\pm$1.75 mV, and the values correspond to energies within the SC gap. One can see the standing waves on the surface along the Se-Se or Fe-Fe directions. Figures \ref{fig1}(e) and \ref{fig1}(f) show the Fourier transformed (FT-) QPI patterns from the Fourier transform operation to Figs. \ref{fig1}(c) and \ref{fig1}(d), respectively. There are three sets of spots with the wave vector lengths of $\boldsymbol{q_1}=2\pi\sqrt{2}/a_{0}, \boldsymbol{q_2}=\pi\sqrt{2}/a_{0}$, and $\boldsymbol{q_3}=2\pi/a_{0}$, consistent with previous reports \cite{20,27}. The scattering spots in the FT-QPI pattern usually originate from the scattering between different Fermi surfaces with similar sizes or different segments along similar directions of a Fermi surface in the momentum space \cite{17}. In FeTe$_{0.55}$Se$_{0.45}$, there is a hole pocket centered at the $\Gamma$ point and four electron pockets at the M points, as illustrated by the schematic diagram in Fig. \ref{fig1}(g). The interpocket scattering wave vectors $\boldsymbol{q_1}$ and $\boldsymbol{q_3}$ connect two electron or two hole pockets, while $\boldsymbol{q_2}$ connects the hole pocket with an electron pocket.

\begin{figure}[tbp]
	\centering
	\includegraphics[width=8.5cm]{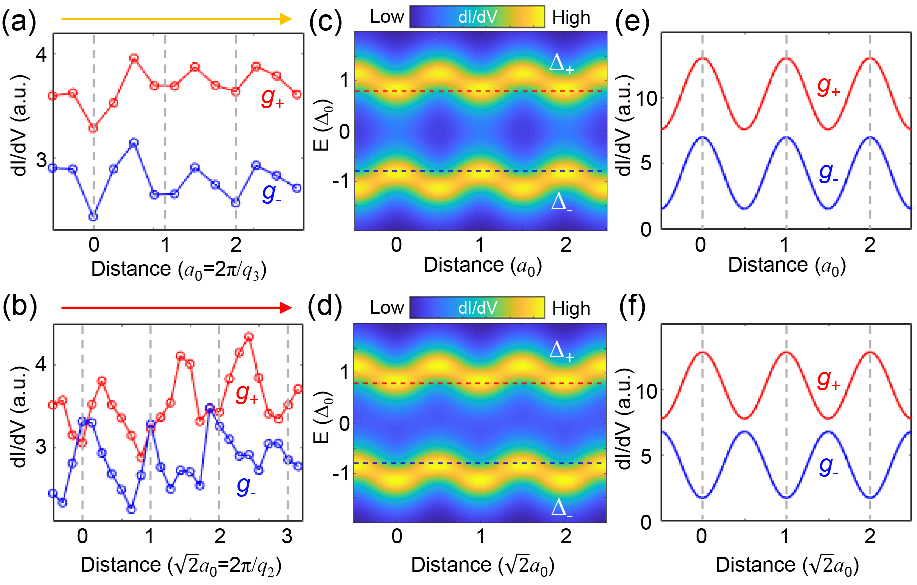}
	\caption{(a),(b) Spatial distributions of dI/dV extracted along the yellow and red lines in Fig.\ref{fig1}(c) or \ref{fig1}(d), respectively. (c),(d) Schematic results based on the PBSI theory of the dI/dV along $\boldsymbol{q_3}$ and $\boldsymbol{q_2}$ directions, respectively. The SC gaps determined from the coherence peaks can be easily identified, as marked by $\Delta_+$ and $\Delta_-$. (e),(f) Distributions of dI/dV at $\pm 0.8 \Delta_0$ extracted from (c),(d), as marked by red and blue dashed lines in (c),(d). The curves are offset for clarity.
	}\label{fig2}
\end{figure}

The widely accepted pairing picture is $s^\pm$ for the iron-based superconductors with both hole and electron pockets \cite{35,36}, which means that the gap signs on the hole and electron pockets are opposite. The phase-referenced (PR-) QPI techniques can be used to judge whether the gap sign changes on the Fermi pocket or between the Fermi pockets \cite{20,21,22,23,24,25,26,27,28}. One of the techniques is based on the defect-bound states (DBS) with multiple impurities \cite{26,27}, and the PR-QPI signal can be expressed as $g_r(\mathbf{q},-E)=|g(\mathbf{q},-E)|\mathrm{cos}(\varphi_{\mathbf{q},-E}-\varphi_{\mathbf{q},+E})$. Here, $\varphi_{\mathbf{q},\pm E}$ are the phases of the QPI signals at $\pm E$. Based on the DBS-QPI technique, $g_r(\mathbf{q},-E)$ will be negative if $\boldsymbol{q}$ connects two Fermi pockets with sign-reversing SC gaps, and positive if the gaps are sign-preserving. Figure \ref{fig1}(h) shows the fourfold symmetrized PR-QPI pattern measured in FeTe$_{0.55}$Se$_{0.45}$. The PR-QPI signal shows negative intensities near the scattering spots around $\boldsymbol{q_2}$, and positive intensities around $\boldsymbol{q_1}$ and $\boldsymbol{q_3}$, providing direct evidence of the sign reversal SC gaps between the hole and electron pockets \cite{27}. The negative value of $g_r(\mathbf{q},-E)$ suggests an anti-phase behavior between the QPI signal $g(\mathbf{r},-E)$ and $g(\mathbf{r},+E)$. We focus on some local regions with stronger scatterings with the real-space periods of $\sqrt{2}a_0$ [along the Fe-Fe direction, near the red line in Fig. \ref{fig1}(c) or \ref{fig1}(d)] and $a_0$ [along the Se-Se direction, near the yellow line in Fig. \ref{fig1}(c) or \ref{fig1}(d)] corresponding to the FT-QPI signal with scattering vectors of $\boldsymbol{q_2}$ and $\boldsymbol{q_3}$, respectively. Figures \ref{fig2}(a) and \ref{fig2}(b) show the spatial dependence of $g(\mathbf{r},\pm E)$ along two arrowed lines across the above-mentioned two kinds of regions selected in Fig. \ref{fig1}(c) or \ref{fig1}(d). One can see that $g(\mathbf{r},-E)$ and $g(\mathbf{r},+E)$ show a clear in-phase feature in Fig. \ref{fig2}(a), while they show an anti-phase feature in Fig. \ref{fig2}(b). These results are consistent with the conclusion obtained from the DBS-QPI technique. 

While conventional QPI analysis focuses on the spatial distribution of LDOS at a fixed energy, it does not account for the spatial variation of the SC gaps (or more precisely, SC coherence peak energies). This raises the question of how quasiparticle scattering influences the spatial distribution of coherence peak energies. According to the recently developed PBSI theory \cite{29}, the scattering vectors associated with sign-preserving (or sign-reversing) gaps can generate anti-phase (or in-phase) modulations of coherence peak energies at positive and negative bias. Specifically, the LDOS contributed by the low-energy eigenstates at positive and negative bias can be expressed as:
$D_+(\omega,\mathbf{r})\sim[1-\cos{(\mathbf{Q}\cdot\mathbf{r}+\phi)}]\delta(\omega-\Delta_S)+[1+\cos{(\mathbf{Q}\cdot\mathbf{r}+\phi)}]\delta(\omega-\Delta_L)$, $D_-(\omega,\mathbf{r})\sim[1-s\cos{(\mathbf{Q}\cdot\mathbf{r}+\phi)}]\delta(\omega+\Delta_S)+[1+s\cos(\mathbf{Q}\cdot\mathbf{r}+\phi)]\delta(\omega+\Delta_L).$
Here, $s = \pm 1$ indicates the relative sign of the gap function at the hot spots, $\mathbf{Q}$ is the scattering vector, and $\Delta_{L,S} = \Delta_0\pm V$ represent the energy levels split by the PBSI effect, where $\Delta_0$ is the average gap and $V$ denotes the magnitude of impurity-scattering potential. We set $V = 0.15\Delta_0$ and the schematic results are shown in Figs. \ref{fig2}(c)-\ref{fig2}(d). The coherence peak positions, marked as $\Delta_+$ and $\Delta_-$, are clearly identifiable. For $s = 1$ (the case of $\boldsymbol{q_3}$), the coherence peaks exhibit particle-hole symmetric (anti-phase) modulations. In contrast, for $s = -1$ (the case of $\boldsymbol{q_2}$), the modulations are particle-hole asymmetric (in-phase). A comparison between Figs. \ref{fig2}(a)-\ref{fig2}(b) and Figs. \ref{fig2}(c)-\ref{fig2}(d) reveals that the phase relation of the LDOS modulations between positive and negative bias is exactly opposite to that of the coherence peak energy modulations. To prevent any confusion, we extract the distributions at $\pm 0.8 \Delta_0$ from Figs. \ref{fig2}(c)-\ref{fig2}(d), as presented in Figs. \ref{fig2}(e)-\ref{fig2}(f). The results show that the schematic distributions agree well with the experiments, demonstrating that the two seemingly contradictory modulation behaviors of LDOS and coherence peak energy are actually consistent and can be explained by the same theoretical framework.

\begin{figure}[tbp]
	\centering
	\includegraphics[width=8cm]{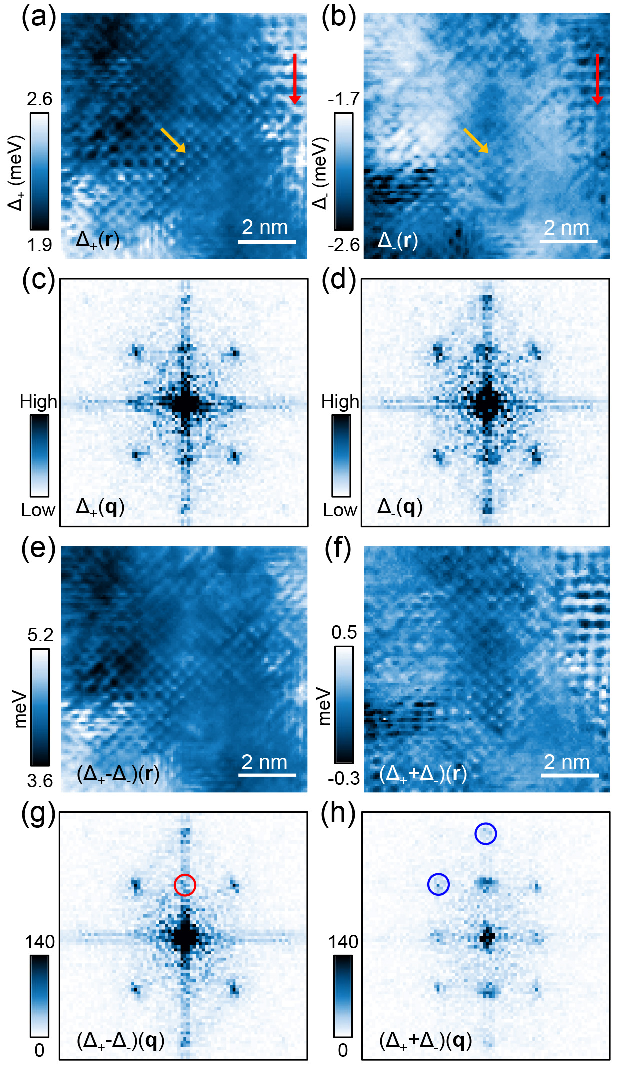}
	\caption{(a),(b) Spatial distributions of the positive ($\Delta_+$) and negative ($\Delta_-$) superconducting gap positions and (c),(d) the corresponding FT results. (e),(f) Spatial distributions of $(\Delta_+ - \Delta_-)$ (usually referred as the SC gap size) and $(\Delta_+ + \Delta_-)$ and (g),(h) the corresponding FT results.
	} \label{fig3}
\end{figure}

In order to measure the evolution of the SC gaps, we divided the region in Fig. \ref{fig1}(a) into a matrix of 112 pixels × 112 pixels and carried out the tunneling spectrum measurements on each pixel. The SC gap can usually be determined by the coherence peak positions. We extract the coherence-peak energies at positive and negative energies of $\Delta_+$ and $\Delta_-$, and the two-dimensional (2D) mappings of them are plotted in Figs. \ref{fig3}(a) and \ref{fig3}(b), respectively. One can see that the coherence-peak energy mappings exhibit real-space modulations similar to those shown in Figs. \ref{fig1}(c) and \ref{fig1}(d). Meanwhile, the FT-patterns in Figs. \ref{fig3}(c) and \ref{fig3}(d) also show scattering spots with wave vectors $\boldsymbol{q_1}$, $\boldsymbol{q_2}$, and $\boldsymbol{q_3}$, which are also similar to the FT-QPI patterns shown in Figs. \ref{fig1}(e) and \ref{fig1}(f). In STM studies, the spectral SC gap value of superconductivity is typically defined as $2\Delta = \Delta_+ - \Delta_-$, and its spatial distribution is shown in Fig. \ref{fig3}(e). Surprisingly, the resulting gap map shows distinct features compared to those of $\Delta_+(\mathbf{r})$ and $\Delta_-(\mathbf{r})$. While the $\boldsymbol{q_2}$-related modulation is most prominent in the upper-right region of Figs. \ref{fig3}(a) and \ref{fig3}(b), it is almost entirely absent in Fig. \ref{fig3}(e). This observation is corroborated by the corresponding FT result in Fig. \ref{fig3}(g), where the $\boldsymbol{q_2}$ component vanishes (highlighted by the red circle), while $\boldsymbol{q_1}$ and $\boldsymbol{q_3}$ remain prominent. For comparison, we also calculated the sum map $(\Delta_+ + \Delta_-)$, as shown in Fig. \ref{fig3}(f). In contrast to the difference map $(\Delta_+ - \Delta_-)$, here the $\boldsymbol{q_2}$-related modulation remains strong, particularly in the upper-right region. The corresponding FT result, as shown in Fig. \ref{fig3}(h), exhibits the opposite trend to Fig. \ref{fig3}(g): the intensities of $\boldsymbol{q_1}$ and $\boldsymbol{q_3}$ are clearly suppressed (though not eliminated) compared to the original $\Delta_+(\mathbf{q})$ or $\Delta_-(\mathbf{q})$, as denoted by the blue circles, while $\boldsymbol{q_2}$ remains robust. These contrasting behaviors suggest distinct phase relations between $\Delta_+(\mathbf{r})$ and $\Delta_-(\mathbf{r})$ for the two sets of wave vectors, ($\boldsymbol{q_1}$, $\boldsymbol{q_3}$) and $\boldsymbol{q_2}$. For $\boldsymbol{q_2}$, the modulations in $\Delta_+(\mathbf{r})$ and $\Delta_-(\mathbf{r})$ are in-phase, leading to the cancellation in the difference map $(\Delta_+ - \Delta_-)$. In contrast, for $\boldsymbol{q_1}$ and $\boldsymbol{q_3}$, the modulations are anti-phase, resulting in their suppression in the sum map $(\Delta_+ + \Delta_-)$. The local anti-phase or in-phase oscillations can also be recognized in the PR analysis, the 2D lock-in analysis \cite{5,6,12,38}, and the Fourier-filtered results of $\Delta_+(\mathbf{r})$ and $\Delta_-(\mathbf{r})$ (see Supplemental Material \cite{37_SM}).

\begin{figure}[tbp]
	\centering
	\includegraphics[width=8.5cm]{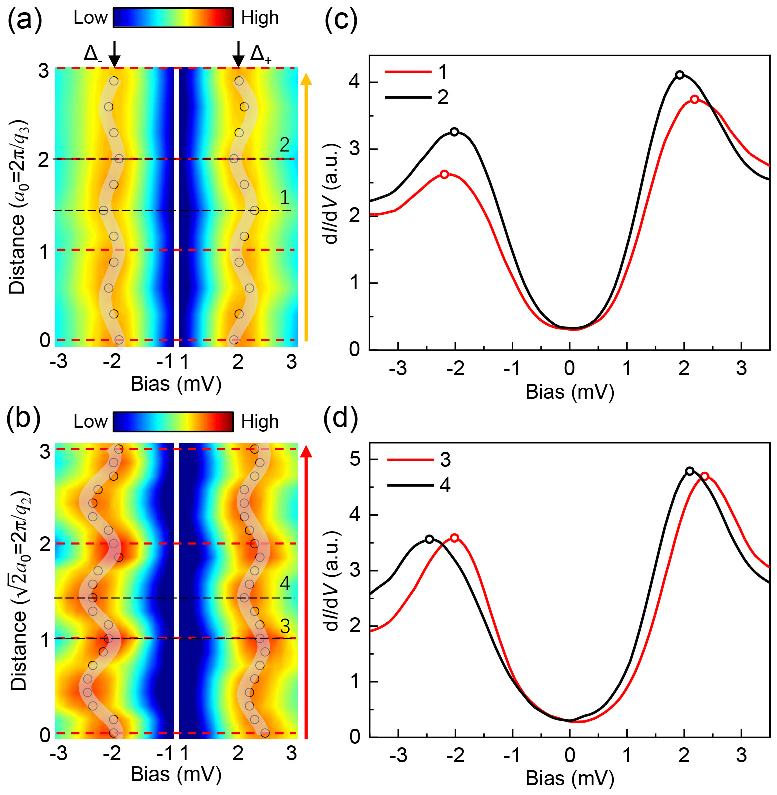}
	\caption{(a),(b) Color plots of tunneling spectra measured along the red (a) and the yellow (b) lines in Fig. \ref{fig3}(a) or \ref{fig3}(b), respectively. The positive and negative sides of the tunneling spectra are normalized by setting the integrated part from 0 to +3.5 mV and from –3.5 to 0 mV to the same value to remove the asymmetric feature of the tunneling spectra. The coherence peak locations are marked by the circles, and the translucent lines with oscillations connecting the circles are guides for the eyes. (c),(d) Typical tunneling spectra with symmetric (c) and asymmetric (d) modulation extracted from (a) and (b) at the positions marked by dashed lines.
	} \label{fig4}
\end{figure}

Similar to the results of Figs. \ref{fig2}(a) and \ref{fig2}(b), we carry out the analysis in the region with a strong modulation amplitude corresponding to the spots near $\boldsymbol{q_2}$ [along the red line in Fig. \ref{fig3}(a) or \ref{fig3}(b)] and $\boldsymbol{q_3}$ [along the yellow line in Fig. \ref{fig3}(a) or \ref{fig3}(b)] and plot the two sets of tunneling spectra in Figs. \ref{fig4}(a) and \ref{fig4}(b), in order to provide a more direct visualization of the two distinct types of SC gap modulations. These spectra have been normalized for better visualization of the SC gap modulations (see Supplemental Material \cite{37_SM}). One can see a clear anti-phase oscillation of the coherence peak energies at positive and negative sides in Fig. \ref{fig4}(a), but an in-phase oscillation in Fig. \ref{fig4}(b). The anti-phase oscillation in Fig. \ref{fig4}(a) results in a spatial modulation of $2\Delta$. This type of spatial modulation of the SC gap is observed in various superconductors with incommensurate \cite{6,12} or (nearly) commensurate \cite{1,2,3,4,5,7,8,9,11,13,14,15} real-space period to $a_0$, and they are regarded as PDW or PDM. In contrast, the in-phase oscillation results in a nearly constant value of $2\Delta$ besides the oscillations of the coherence-peak position at positive and negative sides, similar to the situation observed in overdoped cuprates \cite{39}. In Fig. \ref{fig4}(c), we present two representative tunneling spectra (unnormalized) obtained at the beginning and midpoint of a modulation period from Fig. \ref{fig4}(a), which clearly demonstrate the symmetric characteristic of the modulation: both $|\Delta_+|$ and $|\Delta_-|$ simultaneously reach their maximum or minimum values. This type of modulation naturally persists in the $(\Delta_+ - \Delta_-)$ map, but disappears (or is significantly weakened) in the $(\Delta_+ + \Delta_-)$ map. Similarly, the asymmetric modulation feature in Fig. \ref{fig4}(b) is highlighted by the representative tunneling spectra in Fig. \ref{fig4}(d), where $|\Delta_+|$ is maximum when $|\Delta_-|$ is minimum and vice versa, which clearly differs from Fig. \ref{fig4}(c).

The experimental observations of two types of coherence peak modulations are in good agreement with the recent theoretical picture of PBSI \cite{29}, where the pair breakers should be magnetic \cite{40,41} and nonmagnetic \cite{42,43} impurities for the $\boldsymbol{q_3}$ and $\boldsymbol{q_2}$ scatterings, respectively. Here, the magnetic impurity can be some interstitial Fe atoms \cite{30} under the surface. It should be noted that in our system, the amplitudes of the SC gap modulations are about 5\% and 9\% of the gap average in Fig. \ref{fig4}(a) and Fig. \ref{fig4}(b), respectively, which are much smaller than that in the thin flake samples (up to 30\%) \cite{13}. It was argued that the PDM state may be closely related to the glide-mirror symmetry breaking and nematic distortion in thin flakes \cite{13,16}. However, these two ingredients may not be present in our case.

In summary, we observe two distinct types of SC gap modulations in bulk FeTe$_{0.55}$Se$_{0.45}$. Sign-preserving (reversing) scattering processes give rise to particle-hole symmetric (asymmetric) coherence peak energy modulations, characterized by anti-phase (in-phase) oscillations at opposite energy values, which are consistent with the PBSI theory. Our results offer new insights into the microscopic nature of spatially modulated superconductivity and advance our understanding of the interplay between defect induced QPI and the SC condensate.

This work is supported by the National Key R\&D Program of China (Grants No. 2022YFA1403201, No. 2024YFA1408104), National Natural Science Foundation of China (Grants No. 11927809, No. 12434004). 

$^*$These authors contributed equally to this work.

$^\dag$huanyang@nju.edu.cn

$^\S$hhwen@nju.edu.cn

\end{document}